# A few provoking relations between dark energy, dark matter and pions[1]

Dragan Slavkov Hajdukovic*
PH Division CERN
CH 1211 Geneva 23
dragan.hajdukovic@cern.ch
*On leave from Cetinje, Montenegro

**Abstract**
I present three relations, striking in their simplicity and fundamental appearance. The first one connects the Compton wavelength of a pion and the dark energy density of the Universe; the second one connects the Compton wavelength of a pion and the mass distribution of non-baryonic dark matter in a galaxy; the third one relates the mass of a pion to fundamental physical constants and cosmological parameters. All these relations are in excellent numerical agreement with observations

In this paper I would like to point out three interesting relations of great simplicity and fundamental appearance. The first one connects the Compton wavelength of a pion and the dark energy density of the Universe. As is well known, according to contemporary Cosmology, a mysterious dark energy is responsible for the accelerated expansion of the Universe. One appealing possibility is to identify dark energy with the energy of the physical vacuum; but the trouble (known as the "cosmological constant problem" [1]) is that Quantum Field Theories predict energy density of the vacuum to be many orders of magnitude larger than the observed density of dark energy,

Let me start with the conjecture that a physical vacuum may be modelled as a perfect fluid with temperature $T_v$ determined by:

$$k_B T_v = \frac{1}{2\pi}\frac{\hbar}{c}\ddot{R} \tag{1}$$

Here $k_B$ is the Boltzmann constant; $R$ is the cosmological scale factor and $\ddot{R}$ the corresponding acceleration of the expansion of the Universe. In order to avoid confusion, let's note, that we use cosmological scale factor $R$ as determined by the Friedman equation in the familiar form

$$\frac{kc^2}{H^2 R^2} = \Omega - 1; \quad k = +1, -1, 0 \tag{2}$$

where $\Omega$ is the total energy density of the Universe relative to the critical density, $H$ is the Hubble parameter, and we restrict considerations to the closed universe ( $k = 1$ ) favoured by observations.

In fact Eq. (1) has the same mathematical form as the famous Hawking temperature [2] and Unruh temperature [3], but the key difference is that through Eq.(1) we attribute to the quantum vacuum a universal temperature, having the same value for all observers. If this eventually happened to be true, this conjecture would have major consequences in physics and Cosmology.

Now, there is a surprise. The quotient of energy $k_B T_v$ and the volume $\lambda_\pi^3$ (where $\lambda_\pi \equiv h/m_\pi c$ is the Compton wavelength of a pion) has the same order of magnitude as the dark energy density ( $\rho_{de}$ ), i.e.

$$\rho_{de} = \frac{A}{2\pi}\frac{\hbar}{c}\frac{\ddot{R}}{\lambda_\pi^3} \tag{3}$$

where A is a dimensionless constant.

In the particular case, when dark energy is modelled by the cosmological constant, the choice *A=2*, i.e.

$$\rho_{de} = \frac{1}{\pi}\frac{\hbar}{c}\frac{\ddot{R}}{\lambda_\pi^3} \tag{4}$$

gives for the present-day value $\rho_{de,0} = 6.87 \times 10^{-10} J/m^3$,which is in an excellent agreement with observations [4]. Note that in calculations we have used the Compton wavelength of $\pi^{\pm}$.

---

It remains an open question if Eq. (4) is just a numerical coincidence or if, together with conjecture (1), it has a physical significance.

Let us point out, that nearly half a century ago, an interpretation of the cosmological constant through vacuum energy density was pioneered by Zeldovich [5]. His conjecture

$$\rho_{de} = \frac{c^4}{8\pi G}\Lambda \sim \frac{Gm^6c^4}{\hbar^4} \approx 9.6\times10^{-9}\,J/m^3 \tag{5}$$

(where $\Lambda$ is the cosmological constant and $m$ is close to the mass of a pion) predicts a constant vacuum energy density, while my equation (4) conjectures a dark energy density which is not constant but changes with the expansion of the Universe.

It is amusing that the Compton wavelength of a pion can also be connected with the distribution of the non-baryonic dark matter in a galaxy. Let us remember that all galaxies reside within large halos of dark matter and let us denote by $M_b$ the total baryonic mass of a galaxy, and by $M_{dm}(r)$ the mass of non-baryonic dark matter within a sphere of radius *r*, centred on the centre of the galaxy. The observations suggest that the radial density of dark matter ($\rho_r \equiv dM_{dm}/dr$) has a constant value. The surprise is that, the geometrical mean of the pion mass and baryonic mass of a galaxy, divided by the Compton wavelength of a pion, has the same order of magnitude as the radial dark matter density, i.e.

$$\rho_r \equiv \frac{dM_{dm}}{dr} = \frac{1}{B}\frac{1}{\lambda_\pi}\sqrt{m_\pi M_b} \tag{6}$$

Dimensionless constant *B* may slightly vary from galaxy to galaxy, but in general it is close to the value $B = 2$.

Consequently, Eq.(6) leads to

$$M_{dm}(r) \approx \rho_r(r-a) = \frac{1}{B}\frac{r-a}{\lambda_\pi}\sqrt{m_\pi M_b} \tag{7}$$

where the constant $a$ depends on the galaxy in question..

Just to get an idea about the accuracy of Eq.(6) and Eq.(7) let us consider the case of our Galaxy taking $M_b = 3\times10^{41}kg$ and $a \approx 8kpc$. Then the mass enclosed within 60kpc is about $7.8\times10^{41}kg$ which is in excellent agreement with the observed value $(8\pm1.4)\times10^{41}kg$ in Ref.[6]. The mass within the virial radius (taken to be 250kpc) is about $3.6\times10^{42}kg = 1.8\times10^{12}M_\Theta$, which is once again in good agreement with observations (see [6] and references therein).

My third relation is

$$m_\pi^3 = \frac{\hbar^2}{cG}\frac{H\Omega_\Lambda}{\sqrt{\Omega-1}}\frac{R_0}{R} \tag{8}$$

Here $H$ is the Hubble parameter; index $0$ denotes the present day value of the scale factor $R$; the dimensionless parameters $\Omega$ and $\Omega_\Lambda$ denote respectively the total energy density and dark energy density of the Universe. Hence, the mass of a pion is expressed by both: fundamental physical constants and cosmological parameters.

Note that an incomplete form of relation (8), i.e. proportionality

$$m_\pi^3 \sim \frac{\hbar^2}{cG}H \tag{9}$$

was known to Dirac [7] and Weinberg [8], but there are problems with relation (9). In fact, we are forced to choose $H = H_0$ in relation (9) and even so the left-hand side is about one order of magnitude greater than the right one. In order to save the general form of relation (9) Dirac has suggested that ratio $H/G$ stays constant with time; hence introducing a varying gravitational “constant”. Recently an interesting alternative $H/c = constant$ was studied [9], introducing a varying speed of light. As anyone with a basic knowledge of Cosmology may check, we have completed the old relation (9), in such a way that the mass of a pion does not change with the expansion of the Universe, and as needed, the right-hand side is greater by about one order of magnitude; and it is achieved without invoking varying “constants”.

Of course, Eq.(9) may be completed in a different way than I did, but the key point is that it must be completed; which is fundamentally different from Dirac’s assumption that ratio $H/G$ stays constant with time (i.e. that the gravitational constant $G$ decreases with time). In fact the simplest way to make the right-hand side of Eq. (9) independent of time is to multiply it by $D\sqrt{\Omega_\Lambda}$, where dimensionless constant $D$ must have a numerical value close to $4\pi$ or $\pi^2$. Hence, a plausible alternative to Eq. (8) is

$$m_\pi^3 = 4\pi \frac{\hbar^2}{cG} H \sqrt{\Omega_\Lambda} \quad (10)$$

It is important to note that Eq.(8) and Eq.(10) are valid only if dark energy is modelled by cosmological constant; i.e. a perfect fluid with $w = -1$, in the equation of state $p = w\rho c^2$, connecting pressure and density. More generally, we may drop the assumption that the vacuum energy density is constant and denote density parameter of the vacuum by $\Omega_v$ (instead of $\Omega_\Lambda$). For instance, if dark energy is modelled by a perfect fluid corresponding to $w = -2/3$, the result is

$$m_\pi^3 = \frac{\hbar^2}{cG} \frac{H\Omega_v}{\sqrt{\Omega - 1}} \quad (11)$$

Note that, Eg.(8), and similary Eq.(10) and Eq.(11), may be written as

$$m_\pi^3 = M_P^2 m_x \quad (12)$$

where $M_P = \sqrt{\hbar c/G}$ is the Planck mass, and

$$m_x = \frac{m_\pi^3}{M_P^2} = \frac{\hbar}{c^2} \frac{H\Omega_v}{\sqrt{\Omega - 1}} = 3.25 \times 10^{-68}\, kg \quad (13)$$

The meaning of the mass $m_x$ is not evident, but its value is approximately that of a possible absolute minimum mass in nature [10]. Now, once again there are surprising "coincidences". The first one is that the mass of a neutrino is close to the geometrical mean of the Planck mass $M_P$ and "minimum mass" $m_x$:

$$m_\nu \approx \sqrt{M_P m_x} = 2.66 \times 10^{-38}\, kg \quad (14)$$

If so, the Eq.(12) can be rewritten as

$$m_\pi^3 \approx M_P m_\nu^2 \quad (15)$$

The second surprise is that the Planck mass is close to the geometrical mean of the "minimum mass" $m_x$ and the present-day mass of the Universe ( $M_U$ ).

$$M_P \sim \sqrt{m_x M_U} \quad (16)$$

Too many "coincidences"! My guess is that pions have a still hidden but enormous importance in the Universe. But how is it possible? Pions are just a very tiny fraction of the matter-energy in our world with quarks and leptons as the building blocks. As huge quantities of "real" pions do not exist, it seems natural to me, to suppose that the importance of pions, suggested by the preceding relations is due to "virtual" pions, which are, according to Quantum Field Theories, an inherent part of vacuum fluctuations. But, there are other virtual particle-antiparticle pairs in the vacuum (like, for instance, electron-positron pairs); why are they eventually insignificant compared with pions? Is it too heretical to interpret my relations as a hint that pions somehow dominate the quantum vacuum?